# MODELLING AN AUTOMATIC PROOF GENERATOR FOR FUNCTIONAL DEPENDENCY RULES USING COLORED PETRI NET

Saeid Pashazadeh[1] and Maryam Pashazadeh[2]

[1]Faculty of Electrical and Computer Engineering, University of Tabriz, Tabriz, Iran
pashazadeh@tabrizu.ac.ir
[2]University of Applied Science and Technology of Tabriz Municipality, Tabriz, Iran
maryamPashazadeh@ymail.com

## ABSTRACT

*Database administrators need to compute closure of functional dependencies (FDs) for normalization of database systems and enforcing integrity rules. Colored Petri net (CPN) is a powerful formal method for modelling and verification of various systems. In this paper, we modelled Armstrong's axioms for automatic proof generation of a new FD rule from initial FD rules using CPN. For this purpose, a CPN model of Armstrong's axioms presents and initial FDs considered in the model as initial color set. Then we search required FD in the state space of the model via model checking. If it exists in the state space, then a recursive ML code extracts the proof of this FD rule using further searches in the state space of the model.*

## 1. INTRODUCTION

In design of relational database, normalization is essential step for removing anomalies. Normalization of database requires extracting functional dependencies between attributes of a relation. Deducing new FD rules from existing FD rules are preliminary requirements in database normalization. Armstrong's axioms are used in deducing new FD rules. These axioms are as follows [1]:

```
Augmentation: if A → B, then AC → BC
Transitivity: if A → B and B → C, then A → C
Reflexivity: if B is a subset of A, then A → B
Self-determination: A→ A
Decomposition: if A → BC, then A → B and A → C
Union: if A → B and A → C, then A → BC
Composition: if A → B and C → D, then AC → BD
General Unification: if A → B and C → D, then A∪(C−B)→ BD
```

Automatic database normalization helps database administrator in normalization of database [2].

Colored Petri net is a powerful method for formal verification of various systems [3, 4]. Colored Petri net is extension of classical Petri net and its capability is extended for modelling wide range of systems [5]. It benefits from defining color sets as custom data types and tokens can have different color types. It benefits from using powerful ML language that is an artificial intelligence language and improves extensively functionality of modelling [6].

Various types of proof checkers and proof generators exist. But most of them support proof of logical expressions [7]. Tree proof generator [8], and AProS (Automated Proof Search) [9] are some examples of proof checkers.

In this paper, a model of automatic proof generator is presented using CPN and tested using CPN tool. This model permits us to find a proof of a FD rule based on the initial FD rules using Armstrong's axioms. Next parts of the paper describe the color sets and functions of the model. Then state space analysis of a case study will be discussed.

## 2. COLOUR SETS, INITIAL MARKINGS AND MODEL OF SYSTEM

### 2.1. Colour Sets

Definitions of colour sets that used in modelling of system are as follows:

```
colset ATTRIBUTE = with A | B | C | D| E | F;
colset PRODRULE = with IN|SE|AU|GE|CO|UN|DE|TR;
colset PREDRULELIST = list INT;
colset ATTRIBUTELIST = list ATTRIBUTE;
colset RULEGENERATION = product PRODRULE *PREDRULELIST;
colset FD = record N:INT * F:ATTRIBUTELIST * S: ATTRIBUTE *
G:RULEGENERATION;
colset RULES = list FD;
colset TOKEN = with t;
```

The colour set ATTRIBUTE used to represent the attributes of a relation (table). Colour set PRODRULE is defined to represent name of production rule that is used in the generation of current FDs. Table 1 shows the abbreviations that are used in color set PRODRULE. Colour set PREDRULELIST represent a list that consists the index of FD rules that used in deducting new FD rule using Armstrong's axioms. Its type is the list of integer values. Colour set ATTRIBUTELIST represents a list of attributes of consequent or predecessor part of a FD rule. Its type is a list of ATTRIBUTE.

Table 1.  Abbreviations used to represent the Armstrong's axiom in the model.

| Abbreviation | Description |
|---|---|
| IN | Initial FD |
| SE | Self-determination |
| AU | Augmentation |
| GE | General Unification |
| CO | Composition |
| UN | Union |
| DE | Decomposition |
| TR | Transitivity |

Colour set RULEGENERATION defined to represent the list of FD rules and the index of Armstrong's axiom that used for deduction of new FD rule. Its type is product of PRODRULE and PREDRULELIST. Colour set FD is defined to fully introduce a FD rule. Its type is a record that contains four fields. First field that is denoted with title N, is of type integer, and represent the number (index) of current FD. Second field is denoted with title F is of type ATTRIBUTELIST and represent the list of attributes that constituting the successor part of a FD rule. Third field denoted with the title S is of type ATTRIBUTELIST and represents the list of attributes that constituting the predecessor part of a FD rule. Fourth field denoted with title G is of type RULEGENERATION and represent that which FDs and axiom of Armstrong used for deduction this FD rule. Colour set RULES is defined for representing all FD rules of the database and its color set is of type a list of FD rules. Colour set TOKEN defined as binary valued color set for limiting the concurrent execution of the transitions.

## 2.2. Initial Markings and Variables

Initial markings of the model are as follows:

```
val  InitialAttribs = 1`[A,B,C,D,E,F];
val  InitialRules = [{N=1,F=[A],S=[B,C],G=(IN,[])},
     {N=2,F=[B],S=[E],G=(IN,[])},  {N=3,F=[C,D],  S=[E,F],G=(IN,[])}];
val  FinalFD = {N= 1, F=[A,D],S=[F] ,G=(IN,[])}:FD;
```

Constant InitialAttribs is of colour set ATTRIBUTELIST and is the initial marking of place Attributes and represents the list of attributes that used in the case study. Constant InitialRules is the initial marking of place Rules with colour set RULES and represents the three following initial FD rules of the model:

```
1: A  → BC
2: B  → E
3: CD → EF
```

Constant FinalFD represents the FD rule AD→F that will studied in the following presented case study. Variables of the model are as follows:

```
var L,L1 : RULES;            var al:ATTRIBUTELIST;
var c: BOOL;                 var k: TOKEN;
```

## 2.3. Model of System

Figure 1 shows the CPN model of the system. Places Step1 and Step1D are fusion places.

Although the state space of proposed model contains few distinct nodes but each node contains large number of FD rules. In simple presented case study, number of FD rules reaches up to 1600 FD rules in some states. Increasing the number of database's attributes causes increase of FD rules that can deduce. Permission of running all transitions of the model in nondeterministic form do not have any effect on the results but greatly increases the number of state space nodes and therefore increases the time of generating state space of the model. Limitation on the enabling of transition in a predefined order causes that incoming and outgoing degree of each node in state space graph will decrease. This limitation decreases the size of state space. Transition SelfDetermination fires only once. If any transition cannot produces new FD rules, then no change in the list of FD rules of the place Rules occurs.

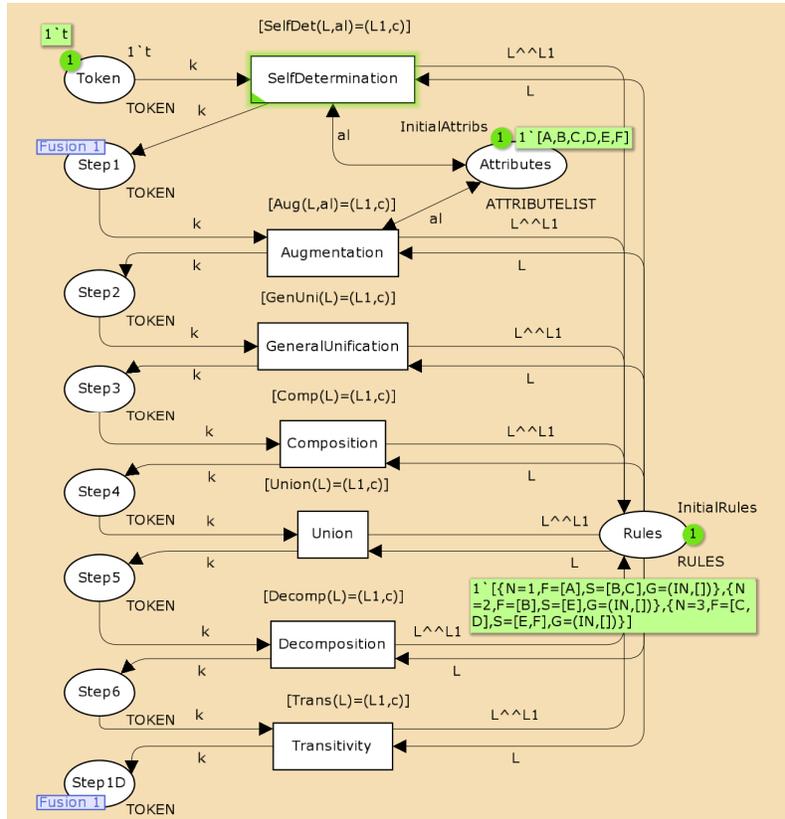

Figure 1. CPN Model of the system.

## 3. PRELIMINARY FUNCTIONS OF MODEL

The system modeled on the assumption that successor and predecessor parts of all FD rules will not contains repeated attribute name. Most functionality of model is base on the model's functions. In this part of the paper, a full description of model's function presents.

Function gerRuleIndex takes a FD rule and a list of FD rules and searches this rule in the list of FD rules. If the rule exists in the list, function returns the position of rule in the list (index starting from zero) in otherwise returns -1 as result. This function calls function isEqual. Antecedent and consequent parts of a rule is a permutation of attributes and this must considered.

```
fun getRuleIndex ( f: FD , (r::L): RULES ) : int =
    if  isEqual(#S f , #S r) andalso isEqual(#F f , #F r)
    then    0
    else
        let val res = getRuleIndex(f,L)
        in   if  (res <> ~1) then
                        res+1
             else        ~1
        end
| getRuleIndex( _,[] ) = ~1;
```

Function isRuleExists takes a FD rule and a list of FD rules as input parameters and searches the existence of this rule in the list of rules. If this rule exists in the list, then function returns true and in otherwise returns false. This function calls the function getRuleIndex to look for position (index) of the FD rule in the list of rules.

```
fun isRuleExists( f: FD , L: RULES ) : bool =
  let  val  n = getRuleIndex(f,L)
  in   if   n <> ~1 then true
       else false
  end
| isRuleExists ( _ , [] ) = false;
```

Function getAttribIndex take an attribute as first parameter and a list of attributes as second parameter. This recursive function returns -1 if the attribute do not exists in the list and in otherwise returns the position of the attribute in the list (starting from index zero).

```
fun getAttribIndex( a : ATTRIBUTE, (r::L):ATTRIBUTELIST ) : int =
  if  a= r then0
  else
     let  val res = getAttribIndex(a, L)
     in   if  ( res <> ~1) then   res +1
          else   ~1
     end
| getAttribIndex( _,[] ) = ~1;
```

Function isAttribExists takes an attribute as first parameter and a list of attributes as second parameter. This function calls the function getAttribIndex and returns true if the result of function getAttribIndex is not -1 then attribute is exist in the list and function isAttribExists returns true and in otherwise the attribute is not exists in the list and function returns false.

```
fun isAttribExists( a : ATTRIBUTE, L: ATTRIBUTELIST ) :
                   bool =
let  val  n = getAttribIndex(a,L)
in   if n <> ~1 then  true
       else   false
end
| isAttribExists ( _ , [] ) = false;
```

Function appendAttrib takes an attribute as first parameter and a list of attributes as second parameter and returns a list of attributes as the result. If the attribute exists in the list then returns the original input list without any changes. In otherwise append the attribute in head of the list and returns new list. This function calls the function isAttribExists and will be used in the function Aug.

```
fun appendAttrib( a : ATTRIBUTE, L: ATTRIBUTELIST ) :
                 ATTRIBUTELIST=
let  val  exist = isAttribExists(a,L)
in   if exist then     L
       else   a::L
end
```

Function difference takes two lists of attributes L1, L2 as input parameters and returns a list of attributes that yields from difference of attributes of list L2 from list L1. This function searches all the attributes of the list L1 in the list L2. Each attribute that is not exists in the list L2 will append to the List L that function will return as the result. This function calls the function isAttribExists and function GenUni called it.

```
fun difference(L1:ATTRIBUTELIST, L2:ATTRIBUTELIST) :
               ATTRIBUTELIST=
  let   val   n1 = List.length(L1)
        val   L = ref  []
        val   i = ref 0
  in
```

```
          while !i< n1   do (
             let val a = List.nth(L1,!i)
             in if not( isAttribExists(a,L2)) then
                     if List.length(!L) = 0 then
                             L :=[a]
                         else L :=  !L ^^ [a]
                else ()
             end;
             i := !i + 1 );         (* while i *)
          !L
      end
| difference( [], _) =[]
| difference( L , []) = L;
```

Function merge takes two lists of attributes L1, L2 as input parameter and returns a list of attributes as result. This function merges the attributes of two lists L1 and L2 by computing the union of attributes of two input lists. This function copies all attributes of list L1 in the resulting list. Then it searches all attributes of the list L2 in the list L1. If any attribute does not exist in the list L1, then function appends it at the tail of the resulting list. In otherwise, for prohibiting of duplicate appearance of an attribute in the result list, the function do not add this attribute to the resulting list. This function calls the function isAttribExists and function GenUni calls it.

```
fun merge(L1:ATTRIBUTELIST,L2:ATTRIBUTELIST):
          ATTRIBUTELIST=
  let   val   n2 = List.length(L2)
        val   L = ref  []
        val   i = ref 0
  in  while !i< n2  do (
          let val a = List.nth(L2,!i)
          in if not( isAttribExists(a,L1)) then
                  if List.length(!L) = 0 then
                          L :=[a]
                     else L :=  !L ^^ [a]
              else ()
              end;
          i := !i + 1 );     (* while i *)
          !L ^^ L1
     end
| merge( [], []) =[]
| merge( L , []) = L
| merge( [], L) = L;
```

Antecedent and consequent parts of a FD rule contains a subset of attributes. Order of attributes in the antecedent and consequent part of a FD rule is not important. Therefore, all permutations of a subset of attributes consider equivalent. Function isEqual takes two subsets of attributes L1, L2 and checks that are they equivalent or not? If they are equal, returns true and in otherwise returns false. If length of lists L1 and L2 are not the same then function returns false. Function searches the existence of all attributes of L1 in the list L2. If one of the attributes of list L1 do not exists in the list L2 then function terminates and returns false.

```
fun isEqual(L1:ATTRIBUTELIST, L2:ATTRIBUTELIST): bool =
  let   val   n1 = List.length(L1)
        val   n2 = List.length(L2)
        val   i = ref 0
        val   j = ref 0
        val   Found = ref true
  in  if  n1 <> n2  then
          false
      else(
         while !i< n1  andalso !Found do (
            let  val   F1 = List.nth(L1,!i)
            in  j := 0;
```

```
                    Found := false;
                    while !j < n2 andalso !Found = false do(
                        let  val F2 = List.nth(L2,!j)
                        in   if F1 = F2 then Found := true
                                else (  )
                        end;
                        j := !j + 1 )   (* while j *)
              end;
              i := !i + 1) ;      (* while i *)
          !Found
          )
   end
| isEqual( [] ,[]) = true
| isEqual( _ , [] ) = false
| isEqual([],_) = false;
```

## 4. FUNCTIONS OF ARMSTRONG'S AXIOMS

All of Armstrong's axioms for deducting new FD rules modeled in the form of separate transitions in the proposed model as in Fig. 1. Guard condition of these transitions plays important rule in the model. All of the guard conditions of transitions relate with the Armstrong's axioms and contains a function that its name is equal with the name of axiom. For simplicity reasons, all of these functions use a common template. In this part, a brief description of this common template for these functions presents. All of the functions present completely in future.

All of the functions Trans, Aug, Decomp, Union, Comp, GenUni, and SelfDet take a list of current existing FD rules as input parameter and return a list of newly generated FD rules using related axiom of Armstrong and a Boolean output value as the results. If a newly generated FD rule exists in the list of current existing FD rules or in the list of newly generated FD rules then it will not append in the list of newly generated FD rules for prohibiting of having repeated FD rules. If a new FD rule does not repeat then function adds this new deducted FD rule in the list of newly generated FD rules. If function cannot produce any new FD rule using related axiom of Armstrong then it returns false and in otherwise it returns true.

Function Trans has the responsibility of generating new FD rules by using current FD rules and based on the transitivity axiom of Armstrong. This function contains two nested loops that both of them start from beginning. This function checks all pairs of current FD rules (only once) and if successor of first rule is equal with predecessor of second rule or wise versa then a new rule can be deducted using transitivity axiom of Armstrong. Newly generated FD rule is a record that contains the index of two FD rules that used in deduction and the new generated FD rule.

```
fun Trans (L: RULES) : RULES * bool =
   let   val  L2 =  ref []
         val  n = List.length(L)
         val  i = ref 0
         val  j = ref 0
         val  cs = ref 0
         val  Found = ref false
         val  Gen = ref false
         val  nr = ref 0
   in
      nr := n+1;
      while !i< n   do (
         let  val    F1 = List.nth(L,!i)
         in  j := 0;
             while !j < n do (
                 if  !i < !j  then
                    let  val F2 = List.nth(L,!j)
                         val ta = { N=(!nr), F = (#F F1),
```

```
                          S=(  #S F2)  ,G=(TR,[#N F1, #N F2])}
                           val tb = { N=(!nr), F = (#F F2),
                          S= (#S F1) ,G=(TR,[#N F1, #N F2])}
                  in  if isEqual(#S F1 , #F F2) then
                           (  cs :=1;
                             Gen := true   )
                      else if isEqual(#S  F2, #F F1)then
                             (  cs := 2;
                               Gen := true )
                          else  ( cs := 0;
                                 Gen := false);
                      case (!cs) of
                          1 => if !Gen=true andalso
                                  not(isRuleExists(ta,L))
                                  andalso
                            not(isRuleExists(ta,!L2))  then
                                ( L2 := !L2 ^^ [ta] ;
                                   nr := !nr +1;
                                   Found := true )
                             else ()
                          | 2 =>if !Gen=true andalso
                           not(isRuleExists(tb,L)) andalso
                            not(isRuleExists(tb,!L2))  then
                                   ( L2 := !L2 ^^ [tb];
                                    nr := !nr +1;
                                    Found := true )
                               else ()
                          |0 => ()
                         end
                     else (  );
                     j := !j + 1 )  (* while j *)
           end;
        i := !i + 1 );        (* while i *)
          (!L2,!Found)
     end
   |  Trans ( [] ) = ([],false);
```

Function Aug takes a list of current FD rules and a list of attributes of a relation as input parameters and returns a list of new FD rules that can deduce from the current FD rules using Armstrong's Augmentation axiom. This function returns a Boolean value as second output parameter. This function chooses all existing FD rules and then combines predecessor and successor parts of each rule separately with all attributes of database. If this newly deductable FD rules are not exists in the list of current FD rules and in the newly deducted FD rules, then append them to the list of newly generated FD rules. If no new FD rules can generate, then function returns false in second output parameter and returns true in otherwise.

```
fun Aug(L: RULES, LA:ATTRIBUTELIST) : RULES * bool=
  let   val   L2 =  ref []
        val   n1 = List.length(L)
        val   n2 = List.length(LA)
        val   i = ref 0
        val   j = ref 0
        val   nr = ref 0
        val   Found = ref false
  in
      nr := n1+1;
      while !i< n1  do (
         let val   F1 = List.nth(L,!i)
         in  j:= 0;
             while !j < n2 do (
               let val  a = List.nth(LA,!j)
                   val t1 = { N= (!nr), F =
                       appendAttrib(a,(#F F1)),
                       S=appendAttrib(a,( #S F1)),
```

```
                        G=(AU,[#N F1])}
        in  if  not(isRuleExists(t1,L)) andalso
               not(isRuleExists(t1,!L2)) then
                  ( L2 := !L2 ^^ [t1];
                    nr := !nr + 1;
                    Found := true)
             else ()
        end;
        j := !j + 1)  (* while j *)
    end;
    i := !i + 1);          (* while i *)
   (!L2,!Found)
 end
|  Aug ( [],_ ) = ([],false);
```

Function Decomp produces new FD rules using decomposition axiom of Armstrong. This function checks all of the current existing FD rules. If successor of a FD rule is contains only one attribute, the function leaves this FD rule and in otherwise generates two new FD rule. Predecessor parts of both new FD rules are same as the predecessor part of old FD rule. Successor part of first new deducted FD rule contains single attribute that is the head attribute in the list of attributed of successor of original FD rule and successor part of second new FD rule is tail of successor of original FD rule. Function searches both of newly deducted FD rules newly generated FD rules in the list of original FDs and before appending them to the list of newly generated FD rules.

```
fun Decomp(L: RULES) : RULES * bool=
  let   val  L2 = ref []
        val  n = List.length(L)
        val  i = ref 0
        val  Found = ref false
        val  nr = ref 0
  in
     nr := n+1;
      while !i< n   do (
        let  val  F1 = List.nth(L,!i)
              val  Len = List.length(#S F1)
        in  if Len > 1 then(
              let  val t1 = {N= (!nr), F= (#F F1) ,
                 S= [ List.hd(#S F1)],G=(DE,[ #N F1])}
              in  if  not(isRuleExists(t1,L)) andalso
                     not(isRuleExists(t1,!L2)) then
                       ( L2 := !L2 ^^ [t1];
                         nr := !nr +1;
                         Found := true )
                  else ();
                  let val t2 = {N= (!nr), F= (#F F1) ,
                       S= List.tl( #S F1) ,
                       G=(DE,[#N F1])}
                  in if  not(isRuleExists(t2,L))
                        andalso
                       not(isRuleExists(t2,!L2)) then
                         ( L2 := !L2 ^^ [t2];
                           nr := !nr +1;
                           Found := true )
                       else ()
                  end
              end
           ) else ()
        end;
        i := !i + 1 );       (* while i *)
     (!L2,!Found)
  end
|  Decomp( []) = ([],false);
```

Function Union generates newly deducted FD rules using union axiom of Armstrong. This function has two nested loops and using them checks all paired combination of current FD rules. Function tests predecessor part of two FD rules by calling function isEqual and if they are equal, then merges their successor parts by calling function merge. Function merge automatically prohibits from appearing an attribute name twice in list of attributes.

```
fun Union(L: RULES) : RULES * bool =
  let   val   L2 =  ref []
        val   n = List.length(L)
        val   i = ref 0
        val   j = ref 0
        val   Found = ref false
        val   Gen = ref false
        val   nr = ref 0
  in
     nr := n+1;
     while !i< n   do (
       let  val   F1 = List.nth(L,!i)
       in  j := 0;
            while !j < n do (
               if  !i < !j  then
                 let  val F2 = List.nth(L,!j)
                      val ta = { N=(!nr) , F = (#F F1),
                                 S=merge(#S F1 ,#S F2),
                                 G=(UN,[#N F1, #N F2])}
                 in   Gen := false;
                      if isEqual(#F F1 , #F F2) then
                         Gen := true
                      else ();
                      if !Gen=true andalso
                         not(isRuleExists(ta,L)) andalso
                         not(isRuleExists(ta,!L2))  then
                           ( L2 := !L2 ^^ [ta] ;
                             nr := !nr + 1;
                             Found := true)
                      else ()
                 end
               else (  );
               j := !j + 1 )  (* while j *)
         end;
         i := !i + 1 );       (* while i *)
       (!L2,!Found)
  end
  |  Union ( [] ) = ([],false);
```

Function Comp produces new FD rules using composition axiom of Armstrong. This function contains two nested loops that both of them start from beginning. This function checks all pairs of current FD rules (only once) and merges the list of attributes of predecessor part of both paired FD rules by calling function merge and considers the new list as the predecessor part of new deducted FD rule. This function merge the successor part of paired FD rules in similar manner and assumes it as the successor of newly deducted FD rule.

```
fun Comp (L: RULES) : RULES * bool =
  let   val   L2 =  ref []
        val   n = List.length(L)
        val   i = ref 0
        val   j = ref 0
        val   Found = ref false
        val   nr = ref 0
  in
     nr := n+1;
     while !i< n   do (
       let  val   F1 = List.nth(L,!i)
```

```
          in j := 0;
              while !j < n do (
                  if  !i < !j  then
                    let  val F2=List.nth(L,!j)
                         val t1={N=(!nr),F=merge(#F F1,#F F2),
                              S=merge(#S F1, #S F2),
                              G=(CO,[#N F1, #N F2])}
                         in  if  not(isRuleExists(t1,L)) andalso
                                 not(isRuleExists(t1,!L2))  then
                                 ( L2 := !L2 ^^ [t1];
                                   Found := true;
                                   nr := !nr + 1 )
                             else ()
                    end
                  else (  );
                  j := !j + 1 )  (* while j *)
            end;
            i := !i + 1 );       (* while i *)
            (!L2,!Found)
      end
   |  Comp ( [] ) = ([],false);
```

Function GenUni produces new FD rules using General Unification axiom of Armstrong. This function contains two nested loops that both of them start from beginning. This function checks all pairs of current FD rules (only once) and by calling functions merge and difference produces two new FD rules using General Unification axiom of Armstrong.

```
fun GenUni(L: RULES) : RULES * bool =
 let  val  L2 =  ref []
      val  n = List.length(L)
      val  i = ref 0
      val  j = ref 0
      val  Found = ref false
      val  nr = ref 0
 in
   nr := n +1;
   while !i< n   do (
      let  val   F1 = List.nth(L,!i)
      in j := 0;
        while !j < n do (
          if  !i < !j  then
            let val  F2 = List.nth(L,!j)
                val Dif21 = difference( #F F2, #S F1)
                val Dif12 = difference( #F F1, #S F2)
                val t1 = { N=(!nr), F = merge(#F F1,
                      Dif21), S= merge(#S F2),
                      G=(GE,[#N F1, #N F2])}
            in if  List.length(Dif21) > 0  then
                  if   not(isRuleExists(t1,L)) andalso
                       not(isRuleExists(t1,!L2)) then
                       ( L2 := !L2 ^^ [t1];
                         nr := !nr +1;
                         Found := true )
                  else ()
                else ();
                if List.length(Dif12) > 0 then
                  let val t2 = {N=(!nr), F= merge(#F F2,
                      Dif12), S=merge(#S F1, #S F2),
                      G=(GE,[#N F2, #N F1])}
                  in if  not(isRuleExists(t2,L))andalso
                         not(isRuleExists(t2,!L2)) then
                       ( L2 := !L2 ^^ [t2];
                         nr := !nr +1;
                         Found := true )
                     else ()
```

```
                   end
               else ()
           end
        else (  );
        j := !j + 1 )   (* while j *)
  end;
   i := !i + 1 );       (* while i *)
 (!L2,!Found)
 end
| GenUni ( [] ) = ([],false);
```

Function SelfDet produces new FD rules using self-determination axiom of Armstrong. This function takes the list of database's attributes as second input parameter and produces new FD rules that predecessor and successor of them are attributes of the database. Function searches all new FD rules in the list of existing FD rules for prohibiting the generation of repeated FD rules.

```
fun SelfDet(L: RULES, LA:ATTRIBUTELIST) : RULES * bool=
let   val  L2 =  ref []
      val  n2 = List.length(LA)
      val  j = ref 0
      val  Found = ref false
      val  nr = ref 0
in    nr := List.length(L)+1;
      while !j < n2 do (
      let val  a = List.nth(LA,!j)
          val  t1 = {N=(!nr), F= [a],S=[a],G=(SE,[])}
      in   if  not(isRuleExists(t1,L)) then
                  ( L2 :=  !L2 ^^ [t1];
                    nr := !nr + 1;
                    Found := true)
              else ()
      end;
      j := !j + 1 );
      (!L2,!Found)
end
| SelfDet ( [],_ ) = ([],false);
```

Figure 2 shows structure chart of model's functions. Functions that are relate to Armstrong's axioms calls the preliminary functions.

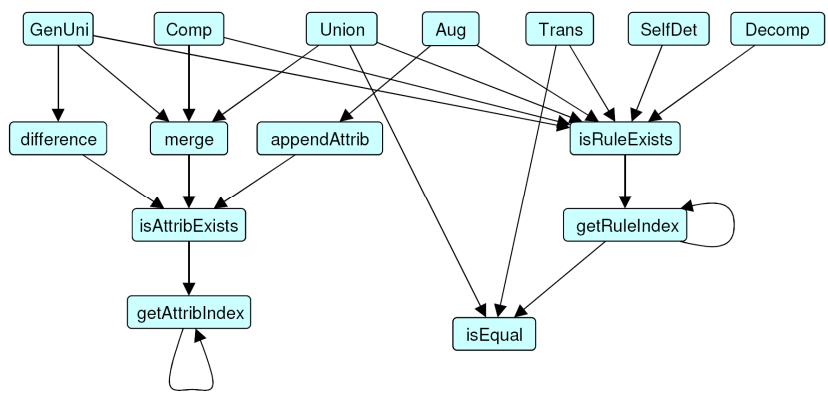

Figure 2.  Structure chart of model's functions.

# 5. STATE SPACE GRAPH OF CASE STUDY MODEL

A sample case study modelled in this paper. Attributes of relation, Initial FDs and the final FD rule that we are looking for its proof are as presented in part 2.2 of paper. Report of state space generation of the model is as follows:

```
State Space
Nodes:  15
Arcs:   15
Secs:   1744
Status: Full
Scc Graph
Nodes:  10
Arcs:   9
```

Figure 3 shows the complete state space graph of the case study model. Extracting proof from state space is a little difficult. It requires model checking of the state space.

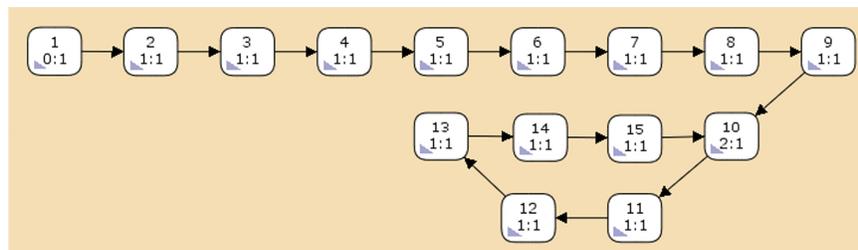

Figure 3.  State space of the case study.

## 6. MODEL CHECKING AND PROOF EXTRACTION

### 6.1. Model checking's functions

Extracting proof of a FD rule requires extensive model checking on the nodes of state space graph. In this part, we explained all required functions that used for extracting proof of a FD rule. Some of these functions have the responsibility of converting information to appropriate string format for producing understandable output. Last functions have the responsibility of searching FD rules that leaded to our final required FD.

Function AtrToStr takes an attribute as input parameter and converts it to equivalent character form that used for producing string output in report generation.

```
fun AtrToStr( atr : ATTRIBUTE ) : string =
  case atr of
       A => "A"
   | B => "B"
   | C => "C"
   | D => "D"
   | E => "E"
   | F => "F"
   | G => "G"
   | H => "H"
   | I => "I"
   | J => "J";
```

Function ALToStr gets a list of attributes and converts it to equivalent string format. This function calls the function AtrToStr.

```
fun ALToStr( L:ATTRIBUTELIST ) : string =
  let  val  n = List.length(L)
        val  i = ref 0
        val  s = ref ""
  in  while !i < n do (
        let  val t1 = List.nth(L , !i )
        in    s := AtrToStr(t1) ^ !s
        end;
        i := !i +1 );
     !s
  end
| ALToStr( [] ) = "";
```

Function PRToStr takes an Armstrong axiom and convert it to equivalent string format.

```
fun PRToStr( pr : PRODRULE ) : string =
  case pr of
    IN => "Initial FD"
  | SE => "Self-determination"
  | AU => "Augmentation"
  | GE => "General Unification"
  | CO => "Composition"
  | UN => "Union"
  | DE => "Decomposition"
  | TR => "Transitivity";
```

Recursive function getRuleIndex2 takes a rule number n as first parameter and a list of FD rules as second parameter and returns the position of rule with rule number n in the list of FD rules (starting from index 0). If no FD rule with number n is exists in the list of rules, then function returns index -1 as the result.

```
fun getRuleIndex2 ( n: INT , (r::L): RULES ) : int =
  if  n = #N r  then  0
  else
      let val res = getRuleIndex2(n,L)
      in  if  (res <> ~1) then
                res+1
          else      ~1
      end
| getRuleIndex2( _,[] ) = ~1;
```

Function FDToStr takes a list of FD rules r and a FD rule fd as input parameters and converts it to equivalent string form. This recursive function calls functions ALToStr, getRuleIndex2, and PRToStr.

```
fun FDToStr( r: RULES, fd : FD ) : string =
  let  val  st = ref ""
        val  s = ref ""
        val  i1 = ref 0
        val  i = ref 0
  in
    if  #1 ( #G  fd ) = IN then
       st :=ALToStr( #F fd )^"-->"^ALToStr(#S fd)
    else (
      let val Len= List.length ( #2( #G fd) )
      in  while !i < Len do (
              i1 := getRuleIndex2( List.nth( (#2
                     ( #G fd)) , !i ) ,r);
              s := !s^ FDToStr( r, List.nth(r, !i1));
              i := !i +1;
              if ( !i < Len ) then  s :=  !s^","
              else ()
```

```
                    );
            st:="{"^!s^ "("^PRToStr( #1( #G fd )) ^") =>
            "^ALToStr(#F fd )^"-->"^ALToStr(#S fd)^"}\n"
      end );
    !st
end;
```

Function ExtractProof takes a list of FD rules as first input parameter and number of a FD rule, then returns the proof in the form of general list, and saves it in the file "Proof.txt" via calling the function FDToStr.

```
fun ExtractProof( r : RULES,  n: INT ) : string =
  let  val  ff = List.nth( r , n)
       val s = ref ""
       val f = TextIO.openOut "Proof.txt"
  in  s := FDToStr( r, ff );
      TextIO.output(f, !s);
      TextIO.closeOut f;
      !s
    end
| ExtractProof ( [] , _) = "";
```

Figure 4 shows the structure chart of functions that are used in extracting proof of a FD rule by analyzing FD rules in nodes of state space graph of the model.

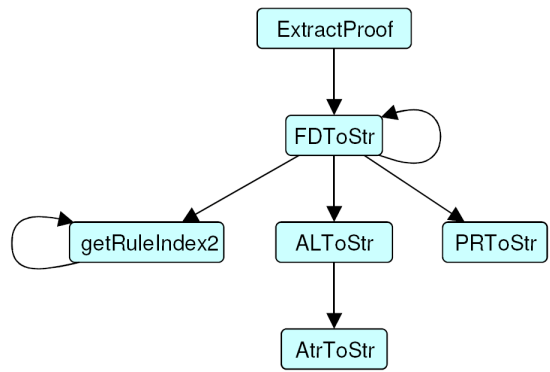

Figure 4. Structure chart of functions that is used in extracting proof of FD rule.

## 6.2. ML Codes of State Space Analysis

Constant FinalFD defines our desired FD rule AD → F. We can use our model to test that, can we deduce this rule using existing FD rules or not?

Function findNodes gets a node n of state space as input and if the desired FD rule (finalFD) appears in the any nodes of the state space of the model then returns true and in otherwise returns false. Function ms_to_col is a build-in function of CPNTool and converts the multi set of state space nodes to a list.

```
fun findNodes n = ( isRuleExists( finalFD ,  ms_to_col
(Mark.FDClosure'Rules 1 n) ) = true);
```
Signiture of function findNodes is as follows:
```
val findNode = fn: Node -> bool
```
Following ML code returns the list of state space nodes that contains our desired FD rule.

```
PredAllNodes    findNodes;
```

It is possible that our desired FD rule appears in more than one nodes of the state space. Output of this ML code in state space graph of case study model is as follows:

```
Val it = [9,8,15,14,13,12,11,10] : Node list
```

For simplicity, we use the first node of the following list for extracting the proof of desired FD rule because this state contains less FD rules. Following ML code extracts index of first node of state space graph that contains our required FD rule:

```
List.hd (PredAllNodes    findNodes)
```

Output of it this ML code is as follows:

```
Val it = 9 : Node
```

Following ML code represents the list of FD rules that deduced in state space node with number 9.

```
ms_to_col (Mark.FDClosure'Rules 1 (List.hd (PredAllNodes    findNodes))
```

Following ML code gets the index of our desired FD rule in the list of FD rules in the first state space node that this FD occurred in it;

```
getRuleIndex( finalFD ,  ms_to_col (Mark.FDClosure'Rules 1 (List.hd
(PredAllNodes    findNodes))));
```

Output of this ML code is as follows:

```
val it = 686 : int
```

Following ML code returns number of FD rules in the list of FD rules in node 9.

```
List.length( ms_to_col (Mark.FDClosure'Rules 1 (List.hd (PredAllNodes
findNodes))));
```

Output of this ML code is as follows:

```
Val it = 1604 : int
```

Following ML code extract the proof of required FD rule (finalFD) from the first node of the state space that this FD rule appeared in it.

```
ExtractProof(  ms_to_col (Mark.FDClosure'Rules 1 (List.hd
(PredAllNodes    findNodes))),
getRuleIndex( finalFD ,  ms_to_col (Mark.FDClosure'Rules 1 (List.hd
(PredAllNodes    findNodes)))));
```

Output of the function ExtractProof is in the form of general lists as follows:

```
{{A-->CB(Augmentation) => AD-->CBD}
,{{B-->E,DC-->FE(General Unification) => BDC-->EF}
(Decomposition) => BDC-->F}
(Transitivity) => AD-->F}
```

For more clarity, we can write the automatically generated proof as following simple form.

```
Augmentation: A → BC ⇒ AD → BCD
General Unification: B → E and CD →EF ⇒ BCD → EF
```

```
Decomposition: BCD → EF ⇒ BCD → F
Transitivity: AD → BCD and BCD → F ⇒ AD → F
```

# 7. CONCLUSION

Automatic proof generation helps database designers in normalization of databases. Colored Petri net is powerful formal method that we can use in modeling and formal verification of wide range of applications. Simple presented model shows that we can use CPN models for automatic proof generation of FD rules. Computing minimal FD rules is under study as an extension to current presented model. Improvements that applied in the model caused that runtime execution of the model decreases greatly but more improvement for decreasing runtime execution time of the model is under study.

**Authors**

**Saeid Pashazadeh** is Assistant Professor of Software Engineering and chair of Information Technology Department at Faculty of Electrical and Computer Engineering in University of Tabriz in Iran. He received his B.Sc. in Computer Engineering from Sharif Technical University of Iran in 1995. He obtained M.Sc. and Ph.D. in Computer Engineering from Iran University of Science and Technology in Iran in 1998, 2010 respectively. He was Lecturer in Faculty of Electrical Engineering in Sahand University of Technology in Iran from 1999 until 2004. His main interest is in the development, modeling and formal verification of distributed systems, and computer security. He is member of IEEE and senior member of IACSIT.

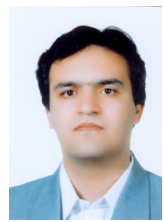

**Maryam Pashazadeh** is Lecturer of University of Applied Science and Technology of Tabriz Municipality in Iran. She works as invited lecturer in Islamic Azad University from 2004. She received her B.Sc. in Applied mathematics from University of Tabriz in 1999. She obtained M.Sc. in Pure Mathematics from Imam Khomeini International University of Qazvin in 2004. Her main interest is modeling and analysis of computer systems, error analysis and performance evaluation of stochastic systems.

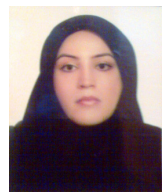